\documentclass{article}

\makeatletter
    \let\@fnsymbol\@alph
\makeatother
\usepackage{caption}
\usepackage{arxiv}

\usepackage[utf8]{inputenc} 
\usepackage[T1]{fontenc}    
\usepackage{hyperref}       
\usepackage{url}            
\usepackage{booktabs}       
\usepackage{amsfonts}       
\usepackage{nicefrac}       
\usepackage{microtype}      
\usepackage{lipsum}		
\usepackage{graphicx}
\usepackage{natbib}
\usepackage{doi}
\usepackage{tabularray}

\usepackage{multirow}
\usepackage{colortbl}
\usepackage{float}

\title{Toward reliable signals decoding for electroencephalogram: A benchmark study to EEGNeX}

\author{
    \href{https://orcid.org/0000-0001-8504-2303}{\hspace{1mm}Xia Chen $^*$ \thanks{Leibniz University Hannover, Germany <xia.chen@iek.uni-hannover.de>}} \\
	\And
	\href{https://orcid.org/0000-0001-5360-4957}{\hspace{1mm}Xiangbin Teng\thanks{Department of Psychology, The Chinese University of Hong Kong, Hong Kong SAR, China}} \\
	\And
	\href{}{\hspace{1mm}Han Chen\thanks{Department of Psychology and Behavioral Sciences, Zhejiang University, China}} \\
	\And
	\href{https://orcid.org/0000-0002-5633-8313}{\hspace{1mm}Yafeng Pan\footnotemark[3]} \\
	\And
	\href{https://orcid.org/0000-0002-0935-4361}{\hspace{1mm}Philipp Geyer\thanks{Leibniz University Hannover, Germany}} \\
}

\renewcommand{\shorttitle}{Reliable signals decoding for electroencephalogram: A benchmark study to EEGNeX}

\hypersetup{
pdftitle={Reliable signals decoding for electroencephalogram: A benchmark study to EEGNeX},
pdfsubject={},
pdfauthor={X Chen},
pdfkeywords={},
}

\begin{document}
\maketitle
\begin{abstract}
    This study examines the efficacy of various neural network (NN) models in interpreting mental constructs via electroencephalogram (EEG) signals. Through the assessment of 16 prevalent NN models and their variants across four brain-computer interface (BCI) paradigms, we gauged their information representation capability. Rooted in comprehensive literature review findings, we proposed EEGNeX, a novel, purely ConvNet-based architecture. We pitted it against both existing cutting-edge strategies and the Mother of All BCI Benchmarks (MOABB) involving 11 distinct EEG motor imagination (MI) classification tasks and revealed that EEGNeX surpasses other state-of-the-art methods. Notably, it shows up to 2.1-8.5\% improvement in the classification accuracy in different scenarios with statistical significance (p < 0.05) compared to its competitors. This study not only provides deeper insights into designing efficient NN models for EEG data but also lays groundwork for future explorations into the relationship between bioelectric brain signals and NN architectures. For the benefit of broader scientific collaboration, we have made all benchmark models, including EEGNeX, publicly available at \href{https://github.com/chenxiachan/EEGNeX}{(https://github.com/chenxiachan/EEGNeX).} 
\end{abstract}


\section{Introduction}

Brain-Computer Interface (BCI) research aims to elucidate communication pathways between the brain and machines \citep{Wolpaw.2000}. From an information theory perspective, modern machine learning models, particularly Neural Networks (NNs) \citep{Schmidhuber.2015}, are evolving to increasingly resemble the human brain’s ability to process, store and communicate information-laden input/output data \citep{Hasson.2020}. Given this context, recent research has leveraged NNs in the BCI domain for information extraction. Among all neuroimaging methods, electroencephalography (EEG) is most commonly used in BCI to its non-invasive nature, low risk, and affordability \citep{Niedermeyer.2005}. NNs excel in receiving and processing EEG signals, which carry implicit information about brain activities, without prior knowledge, thereby contributing to our understanding of the human brain.

As multi-channel time-series data, EEG signals carry implicit information of brain activities. The decoding and representation of these signals are crucial in EEG data processing. In this regard, we observe that two primary types of NNs are adopted in the EEG community: Recurrent Neural Networks \citep{Ruffini.2016,Craik.2019}(RNNs), primarily referring to Long Short-Term Memory, LSTM \citep{Hochreiter.1997} and Gated Recurrent Unit, GRU \citep{Chung.2014}, which are effective in uncovering the hidden state correlations from sequential data, and Convolutional Neural Networks (Conv. or CNNs) \citep{Alzubaidi.2021,Altaheri.2021,Lawhern.2018}, which are proficient at extracting information from spatial correlations.

Despite established research demonstrating the performance advantages of various NN designs in decoding EEG signals through implicit information extraction and end-to-end representation learning \citep{Craik.2019,Hosseini.2020}, a research gap persists: Most studies focus primarily on boosting classification accuracy on a specific dataset or task type. This is typically achieved through manual, explicit feature extraction via domain knowledge preprocessing \citep{Craik.2019}, regularization \citep{Deng.2021}, and transformation techniques \citep{Huang.2020,Altaheri.2021}, along with fine-tuned hyperparameters. As a result, understanding the generalization performance of representation learning abilities across different NNs becomes challenging. This study addresses this issue by examining the performance of previous approaches under the same conditions to guide further improvement. In other words, we seek to answer the key question:

\begin{itemize}
    \item \textit{To what extent do various neural network designs contribute to the efficient representation learning of raw EEG data?}
\end{itemize}

To address this question, we make contributions in this study in three-fold:  

\begin{enumerate}
    \item \textbf{Thorough Evaluation of NN Performance}: We thoroughly evaluated the performance of 16 different fundamental NNs models in four EEG-based BCI classification tasks. The results revealed the efficacy of different NNs in representing EEG-based information and provided insights into designing efficient NN models for EEG-based classification tasks.
    \item \textbf{Development of a Novel NN Architecture}: Based on the literature review, benchmark test results and successful key component implementation of advanced NNs, we provide a trajectory from an advanced NN architecture, EEGNet, to EEGNeX. We intensively reinforce the model’s spatial, temporal, and cross-channel global representation learning efficacy. Compared to typical algorithm pipelines evaluated in Mother of All BCI Benchmarks (MOABB) \citep{Jayaram.2018}, EEGNeX shows higher accuracy with statistical significance on 11 diverse EEG motor imagination (MI) datasets.
    \item \textbf{Open-Source Testbed for Research}: We open-source all models and EEGNeX to provide a testbed for research and to reduce the difficulties of NN implementation in the BCI domain.
\end{enumerate}

\begin{figure}[h]
	\centering
	\includegraphics[width=17cm]{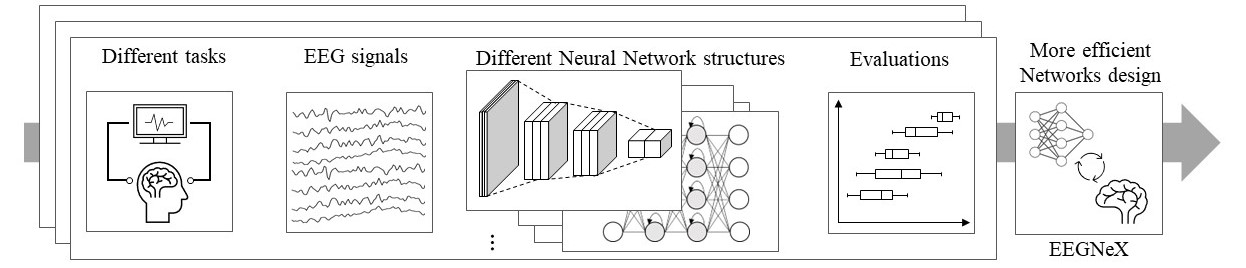}
	\caption{Overall visualization of the research pipeline; We implement different NN architects into variate EEG-based, BCI classification tasks in open-source. Based on the benchmark results, we aim to alleviate the difficulty of finding the path to adapt NNs efficiently toward higher accuracy in the domain. EEGNeX, a novel NN structure is proposed with validation in this study.}
	\label{fig:1.jpg}
\end{figure}

The remainder of this paper is organized as follows: Section \ref{Sec2} reviews the development of neural networks in the domain. The design of benchmark models, target dataset selection, training strategy, and benchmark insights are described in Section \ref{Sec3}, followed by an analysis with a proposal of serial structural innovations based on EEGNet with a comprehensive architecture description of EEGNeX (Section \ref{Sec4}). The accuracy novelty of EEGNeX is validated in Section \ref{Sec5} with the discussion of the future direction in Section \ref{Sec6}. Section \ref{Sec7} concludes the study. Figure \ref{fig:1.jpg} illustratively summarizes the objective of this study.

\section{Literature review: Development of NNs in EEG-based classification tasks}\label{Sec2}

This section provides a concise review of two types of Neural Network (NN) adaptations and their evolution in EEG-based BCI classification tasks, alongside a parallel comparison of NN development in their original domains.

The first well-known NN adaptation in EEG classification was ConvNet \citep{Cecotti.2010}. It consists of two convolutional layers for extracting temporal and spatial features, followed by dense layers, and applied to a P300 task. A similar CNN-based approach supplemented with max-pooling layers \citep{Schirrmeister.2017} showed improved performance compared to a well-known, domain-knowledge-based algorithm - filterbank common spatial patterns (FBCSP) \citep{Chin.2009}. EEGNet builds upon previous methods and combines depthwise convolution with separable convolution \citep{Lawhern.2018} to enhance cross-channel spatial feature extraction. The integration of a dropout layer, batch normalization layers, kernel constraint, and max-pooling stride made EEGNet more efficient in terms of parameters use and cross-section performance compared to ShallowConvNet (two convolution layers) and DeepConvNet (five convolution layers). 

Several extended EEGNet variants incorporating domain knowledge have been proposed, including EEG-inception \citep{Riyad.2020}, which uses multi-head convolutions as an inception block with different kernel sizes (receptive fields) and pointwise convolution; FBCNet \citep{Mane.2021}, which presents a hybrid approach combining bandpass filtering and different narrow filter banks to improve performance on MI tasks; EEG-TCNet \citep{Ingolfsson.2020}, which adds a temporal convolutional network (TCN) \citep{Bai.2018} to strengthen the temporal feature extraction; and EEG-ITNet \citep{Salami.2022}, which combines the inception block and TC block, and claimed accuracy advance.

In terms of RNN-based models, less attention has been dedicated to their application in EEG-based classification tasks \citep{Craik.2019}. A two-layer LSTM implementation has been utilized for emotion recognition and proven to outperform statistical-based feature selection methods with traditional machine learning classifiers \citep{Alhagry.2017}. Similarly, LSTM implementations with feature extractors in MI tasks have reported superior performance compared to conventional methods and other deep networks, including CNNs \citep{Wang.2018}. Another RNN variant, GRUs, has been coupled with the attention mechanism \citep{Vaswani.2017} and applied in emotion classification tasks \citep{Chen.2019}. This model outperforms both CNNs and LSTMs. Subsequently, a structure combining CNNs and LSTMs was proposed for the same task, offering faster training times. In this structure, CNNs handle the spatial information from EEG signals, while RNNs extract the temporal information \citep{Wilaiprasitporn.2019}. 

Originally, CNNs and RNNs were primarily developed as traditional models for tasks in different domains, such as speech recognition (Natural Language Processing, NLP, temporal) and computer vision (CV, spatial), respectively. The intersection between these two models emerged with the introduction of the attention-based RNN Transformer, BERT \citep{Devlin.2018}, proposed for language translation, and ViT \citep{Dosovitskiy.2020} for image recognition. Although ViT outperforms traditional CNN architectures such as VGGNet \citep{Simonyan.2014}, ResNet \citep{He.2016}, and EfficientNet \citep{Tan.2019}, the advantages of such hybrid NN architectures continue to be a topic of ongoing discussion: for instance, ConvNeXt \citep{Liu.2022} fine-tuned the ResNet architecture to further improve accuracy and scalability, reaching the level of Transformers while preserving the simplicity and efficiency of standard convolution networks. Given that BCI seeks to interpret implicit information from brain activities, requiring both temporal and spatial decoding, it is crucial to explore the information representation processes for different NNs, learning from their development trajectory and adapting to the BCI domain.

\section{Preliminary study: Benchmark test}\label{Sec3}

This section adopts a reductionist approach, starting with a benchmark test. Its objective is to investigate the performance of various fundamental NN layer types and their derivatives in the extraction of implicit information. The goal is to attain a quantitative comprehension of the principal NN types and garner insights into designing more efficient networks for EEG-based classification tasks.

\subsection{Candidate methods}

We focus on implementing performance comparison among two main types of NN structures: CNNs and RNNs, with their variations and combinations. Four basic layer types and their variations in TensorFlow \citep{abadi.2016} were chosen for investigation: 2D convolution layer (\textbf{Conv2D}), 1D convolution layer (\textbf{Conv1D}), \textbf{LSTM} layer, \textbf{GRU} layer, and their combinations. 

Moreover, several trending CNN variations whose mechanisms might affect the representation learning ability of NNs are chosen, including:
\begin{itemize}
    \item \textbf{Depthwise separable convolutions} \citep{Chollet.2017}: consist of first performing a depthwise spatial convolution (which acts on each input channel separately) followed by a pointwise convolution that mixes the resulting output channels.
    \item \textbf{Causal padding}: initially designed in WaveNet \citep{vandenOord.2016} for sequential sound generation, as an implementation option in the CNN layer. It pads the layer's input with zeros in the front so values of early time steps can be predicted.
    \item \textbf{Dilation}: add the convolution layer with spaces in-between to support the exponential expansion of the receptive field without loss of resolution or coverage. It aggregates multi-scale contextual information and is firstly proposed for image semantic segmentation \citep{Yu.2015}.
\end{itemize}

Following the throughout reviews of EEG-based BCI classification tasks \citep{Craik.2019,Hosseini.2020}, we reproduced common networks and adopted well-proved regulation procedures in corresponding NNs:

\begin{itemize}
    \item Using kernel regularizers in RNNs.
    \item Using a CNN layer and batch normalization \citep{Ioffe.2015} with e exponential linear unit (ELU) activation function \citep{Shah.2016} as a standard CNN component.
    \item Adding average/max pooling layers at the final layer of CNN to account for local translation invariance.
\end{itemize}

A reproduction of EEGNet is integrated into the benchmark. In this study, we chose EEGNet-8,2, because of its better performance than EEGNet-4,2 (8 and 4 stand for different filter numbers of the first convolution layer in EEGNet), based on results from the original paper \citep{Lawhern.2018}. In this study, 16 NNs are chosen as candidates for the benchmark. Their macro design and detailed architectures of all NN candidates are available in Appendix. 

\subsection{Datasets}

To conduct a comprehensive and in-depth investigation of EEG representation learning and implicit information extraction capabilities on different NNs, we tested on four diverse EEG datasets with the consideration of covering within/cross-subjects, different tasks, paradigms, trial lengths, data sizes, and channels number. A characteristic overview of all datasets is presented in Table \ref{tab:table1}, detailed descriptions and preprocessing procedures is available in Appendix.

\begin{table}
\small
\centering
\SetTblrInner{rowsep=0pt}
\begin{tblr}{
  width = \linewidth,
  colspec = {Q[90]Q[290]Q[77]Q[56]Q[69]Q[83]Q[140]Q[127]},
  column{3} = {c},
  column{4} = {c},
  column{5} = {c},
  column{6} = {c},
  column{7} = {c},
  column{8} = {c},
  hline{1-2,6} = {-}{},
}
\textbf{Paradigm} & \textbf{Dataset} & \textbf{Subjects} & \textbf{Trials} & \textbf{Classes} & \textbf{Channels} & \textbf{Class Imbalance?} & \textbf{Size}\\
ERP & MNIST of BRAIN DIGITS (MBD) & 1 & 224 & 10 & 14 & No & (64302, 14, 224)\\
MI & OpenMBI Data (MBI) & 54 & 200 & 2 & 20 & No & (10800, 20, 512)\\
SMR & BCIC-IV-2A Data (SMR) & 9 & 500 & 4 & 22 & No & (2592, 22, 500)\\
ERN & Feedback Error-Related
  Negativity (ERN) & 26 & 160 & 2 & 56 & Yes, \textasciitilde{}3.4:1 & (5440,
  56, 160)
\end{tblr}
\caption{Description of experiment dataset collections. Class imbalance, if present, is given as odds, means different classes have uneven odds, which is given as the average class imbalance over all subjects.}
\label{tab:table1}
\end{table}

Particularly, two out of four chosen datasets (SMR and ERN) are the same as in the original EEGNet study \citep{Lawhern.2018} for comparative evaluation. The SMR dataset is also used in the performance evaluation of EEG-inception, EEG-TCNet and EEG-ITNet studies. The evaluation was conducted within-subject, as this minimizes the effect of non-stationarity. It is worth mentioning that some datasets in MOABB have small sample sets of data within-session, which increases the difficulties of training NN algorithms. The accuracy metric scores are determined using 5-fold cross-validation to validate the general information extraction and representation performance.

\subsection{Training strategy}

All datasets are grouped according to events (records of all channels for one event) and tested with candidate models for ten rounds under the same train/validation/test ratio of 0.75/0.125/0.125 (training data size equivalent to a 4-fold cross-validation test) with random split seeds each round. As the sign-to-noise ratio of EEG signals is relatively low, NNs must avoid memorizing the noise components from EEG signals to end up overfitting the dataset. 

To ensure that different modeling structures present their best performance, we train the model in a self-monitor behavior for each dataset to exclude the potential impact of different feature engineering processes or data size on the model performance. All candidate methods were run under the same two-stage scheme: 
\begin{enumerate}
    \item The training set is processed under a batch size of 128 for each epoch with a default learning rate of 0.001. The accuracy of the trained model is monitored on the validation set, and the learning rate is reduced by half if accuracy stops improving for five epochs. The training process stops when there is no improvement in the validation set for 20 epochs to prevent overfitting.
    \item The model performance is evaluated by running on the test set to avoid data leakage and represent the performance of all benchmark models in the real-world production environment.
\end{enumerate}

\subsection{Performance evaluation and insights}

Table \ref{tab:table2} presents the performance result of exhaustive evaluation on four datasets across all candidate models. 

\begin{table}
\small
$$\vbox{
\offinterlineskip
\halign{
\strut
\vrule height1ex depth1ex width0px #
&\kern3pt #\hfil\kern3pt
&\kern3pt \hfil #\hfil\kern3pt
&\kern3pt \hfil #\hfil\kern3pt
&\kern3pt \hfil #\hfil\kern3pt
&\kern3pt \hfil #\hfil\kern3pt
&\kern3pt \hfil #\hfil\kern3pt
&\kern3pt \hfil #\hfil\kern3pt
&\kern3pt \hfil #\hfil\kern3pt
&\kern3pt \hfil #\hfil\kern3pt
\cr
\noalign{\hrule}
 &  {\bf Dataset}  & \multispan2\kern3pt \hfill  {\bf MBD} \hfill\kern3pt  & \multispan2\kern3pt \hfill  {\bf MBI} \hfill\kern3pt  & \multispan2\kern3pt \hfill  {\bf SMR} \hfill\kern3pt  & \multispan2\kern3pt \hfill  {\bf ERN} \hfill\kern3pt \cr
\noalign{\hrule}
 &  {\bf Model Name}  & Accuracy & Deviation & Accuracy & Deviation & Accuracy & Deviation & Accuracy & Deviation\cr
\noalign{\hrule}
 & Single\_LSTM & 13.31 & 0.49 & 53.78 & 0.79 & 27.38 & 2.59 & 71.43 & 0.53\cr
 & Single\_GRU & 14.49 & 0.49 & 53.80 & 1.81 & 27.25 & 2.66 & 71.01 & 0.42\cr
\noalign{\hrule}
 & Conv1D & 15.45 & 0.42 & 68.22 & 0.72 & 30.80 & 2.23 & 71.03 & 0.85\cr
 & Conv1D\_Dilated & 15.11 & 0.28 & 69.87 & 0.69 & 31.64 & 3.06 & 70.31 & 0.74\cr
 & Conv1D\_Causal & 15.11 & 0.44 & 67.71 & 0.96 & 30.62 & 2.91 & 69.31 & 0.35\cr
 & Conv1D\_CausalDilated & 15.44 & 0.29 & 67.91 & 0.68 & 30.49 & 1.64 & 69.47 & 0.57\cr
\noalign{\hrule}
 & Conv2D & 15.20 & 0.38 & 67.09 & 0.40 & 32.99 & 1.75 & 69.65 & 0.12\cr
 & Conv2D\_Dilated & 15.15 & 0.29 & 68.76 & 0.81 & 33.46 & 1.96 & 68.88 & 0.28\cr
 & Conv2D\_Separable & 16.26 & 0.45 & 68.67 & 0.90 & 34.72 & 3.13 & 72.26 & 0.10\cr
 & Conv2D\_Depthwise & 16.20 & 0.28 & 68.90 & 0.75 & 36.51 & 2.27 & 71.54 & 0.43\cr
\noalign{\hrule}
 & Conv\_LSTM2D & 16.18 & 0.34 & 69.74 & 0.73 & 36.61 & 1.91 & 71.13 & 0.11\cr
 & Conv\_GRU2D & 16.14 & 0.32 & 68.25 & 0.77 & 36.70 & 1.94 & 70.95 & 0.16\cr
 & Conv\_LSTM1D & 15.77 & 0.34 & 69.57 & 0.80 & 30.77 & 1.63 & 69.90 & 0.42\cr
 & Conv\_GRU1D & 15.75 & 0.31 & 67.48 & 0.80 & 29.41 & 1.68 & 71.85 & 0.19\cr
\noalign{\hrule}
 & EEGNet-8,2 &  {\bf 16.69}  & 0.29 &  {\bf 70.81}  & 0.91 &  {\bf 56.79}  &  2.06{\bf }  &  {\bf 72.13}  & 0.37\cr
 & EEGNet-8,2* & 16.53 & 0.31 & 70.19 & 0.98 & 57.14 & 1.99 & 71.45 & 0.34\cr
\noalign{\hrule}
}
}$$
\caption{Performance result of exhaustive evaluation on four datasets across all candidate models. The average accuracy and deviation are calculated based on ten rounds of the train-validation process with random seeds in the data split. In addition, model parameter sizes are presented along to represent the network complexity and search space size. To validate our implemented EEGNet performance, the Tensorflow\protect\footnotemark[1] version of EEGNet (with mark *) is included}
\label{tab:table2}
\end{table}
\footnotetext[1]{https://github.com/vlawhern/arl-eegmodels}

According to the accuracy results across four datasets, some primitive findings regarding different NN architectures’ performance are collected and categorized in three aspects:

\begin{enumerate}
    \item Sequential processing methods such as Recurrent Neural Networks (RNNs) and one-dimensional Convolutional Neural Networks (Conv1D) with a causal structure do not effectively extract and represent implicit EEG information. No significant performance differences were observed between Long Short-Term Memory (LSTM) and Gated Recurrent Unit (GRU) models. CNN-based models on average outperformed RNNs in classification accuracy. Since all EEG datasets comprise three-dimensional, multi-channel time-series data (sample, channels, time/trials), the \textbf{extraction of cross-channel spatial information} (in the channel dimension) \textbf{is more efficient than the extraction of temporal information} (in the trial/time dimension).
    \item Within the fundamental structure of CNNs, two-dimensional convolution (Conv2D) generally achieved higher accuracy than Conv1D, as it does not need the compression of multi-channel EEG signals into a sequential format and it learns patterns across channels. Among the various variations, the separable 2D convolution architecture shows the best ability for learning temporal summaries \citep{Lawhern.2018}. The stability of accuracy improves when depthwise-separable 2D convolution is combined with EEGNet, with the key factor being the \textbf{mechanism of depthwise convolution that supports efficient cross-channel spatial feature learning.}
    \item The performance of hybrid model structures (CNN-LSTM, CNN-GRU) exhibited considerable accuracy improvements compared to any single original model. Notably, the results of hybrid models with Conv1D layers in all datasets were less accurate than those consisting of three stacked layers of Conv2D (see Appendix). This suggests that \textbf{extracting additional temporal information after appropriate spatial filter/learning is beneficial for further improving model performance}.
\end{enumerate}

\begin{figure}[h]
	\centering
	\includegraphics[width=10cm]{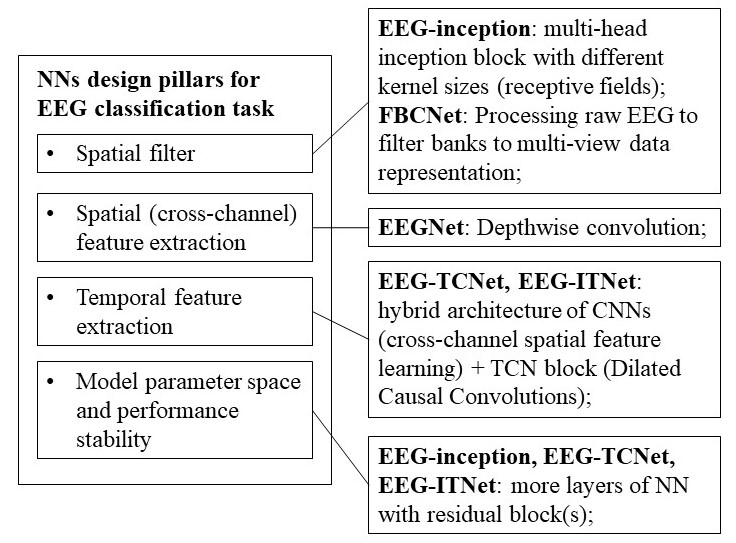}
	\caption{Network architecture design pillars for efficient EEG-based BCI classification tasks with proven models.}
	\label{fig:2.jpg}
\end{figure}

In light of our experimental results and literature review findings, we identify four key areas that contribute to the design of NN architecture for efficient EEG-based BCI classification tasks: spatial filtering, spatial (cross-channel) feature learning ability, temporal feature learning ability, and model parameter space with performance stability. We found that similar existing NNs all benefited in performance improvement by strengthening at least one pillar. Figure \ref{fig:2.jpg} presents these key pillars and corresponding NN designs.

\section{Roadmap: from EEGNet to EEGNeX}\label{Sec4}

In the previous experiment, EEGNet presents the best performance among all candidate models, as: 1) The first part – the 2D convolution layer extracts the spectral representation of EEG input; 2) The second part – the kernel size of depthwise 2D convolution is tailored to the number of channels and directed to perform convolutions across channels. 3) The third part – separable 2D convolution conducted as information extraction and learning. By revisiting the three parts mentioned above and adapting to modern network component designs, we can modify the EEGNet toward higher general performance.

In this section, we provide a trajectory from EEGNet to EEGNeX, incorporating findings from the previous section as well as key component implementations from ConvNeXt \citep{Liu.2022}. The roadmap starts from a standard EEGNet and progresses with adjustments to the network architecture in different parts. Sequentially, we test changes at the micro-level within the fixed network structure. We set the EEGNet as the backbone and tracked the successful steps tested on datasets described in Section 3 with proven performance improvement by ten rounds of validation. 

\begin{figure}[h]
	\centering
	\includegraphics[width=17cm]{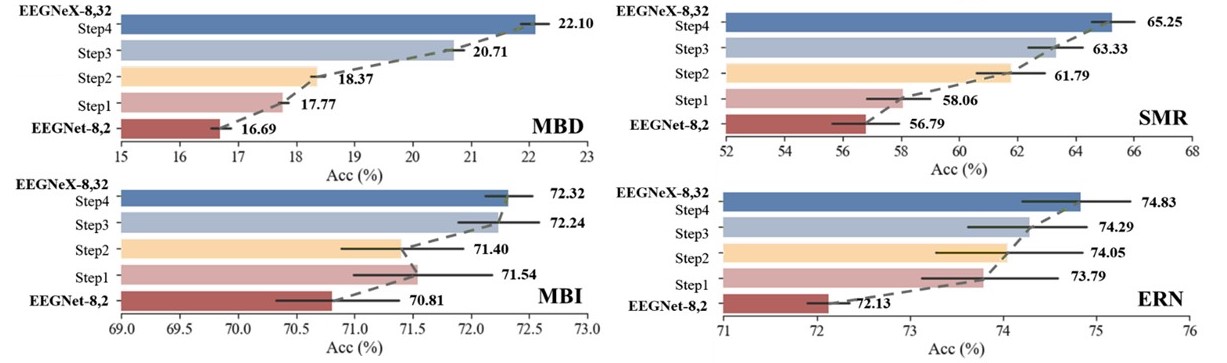}
	\caption{The roadmap of modifying from EEGNet - 8,2 toward EEGNeX - 8,32. The bars record the ten times average performance of the model at each step. The black horizontal lines represent the deviation to reflect model performance stability. In the end, the pure convolution-based network structure, namely EEGNeX, outperforms the EEGNet – 8,2 on all datasets.}
	\label{fig:3.jpg}
\end{figure}

The summarized roadmap is as follows four-steps: 1) Reinforce the spatial representation extraction from EEG input, 2) Replace the separable convolution with two 2D convolutions in the general architecture, 3) Inverse bottleneck structure, 4) Increase the receptive field of the layers with dilation and fewer activations. The modification purposes and detailed procedure with the results of each step are described as follows. The quantitative results are presented in Figure \ref{fig:3.jpg}.

\textbf{Step 1. Thicken the first part of spatial information extraction}

Based on the study from Section \ref{Sec2}, we noticed that the current structure of EEGNet is designed in three parts with the particular purpose of information processing. The first part that aims to extract spectral information from the EEG input and is shallow, with only one Conv2D block with a kernel size of 32 and output filters of 8. Based on the previous study \citep{Schirrmeister.2017} and considering that the general network layer depth should not be too deep, we duplicate one more Conv2D block in the first part. It is worth mentioning that the adjustment we made is the most intuitive solution. The key idea is to strengthen the network spatial information extraction ability, and a more optimal design in this part is likely to exist (see Figure \ref{fig:2.jpg}). 

\textbf{Step 2. Replace the separable convolution with two 2D convolutions}

\begin{figure}[t]
	\centering
	\includegraphics[width=16cm]{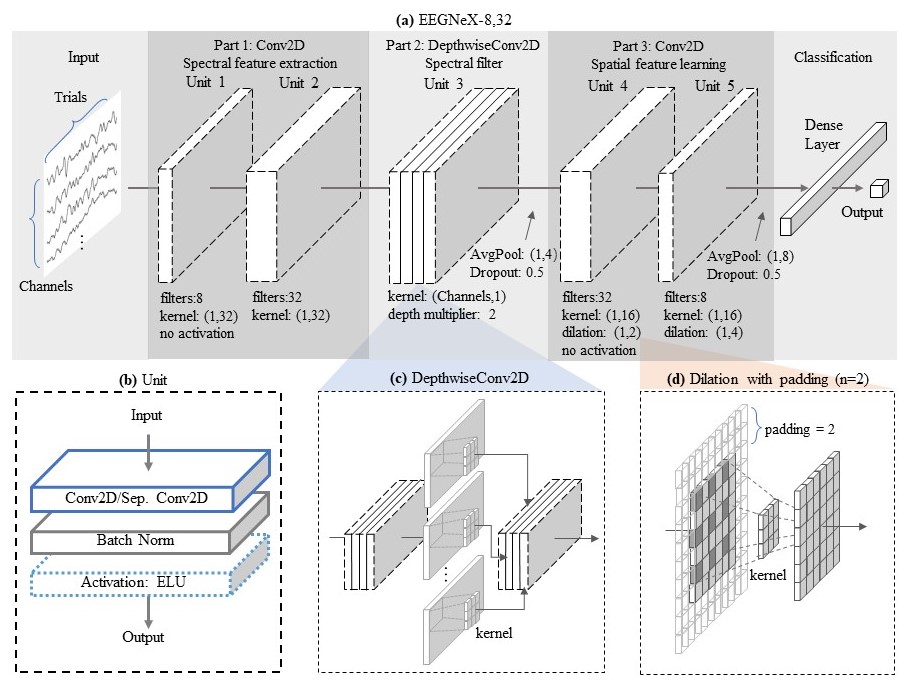}
	\caption{EEGNeX architecture; (a) the general structure visualization of EEGNeX-8,32. An EEGNeX is composed of five units. 8 means the number of temporal filters, while 32 stands for the kernel size; (b) The structure of the unit, which consists of a convolution layer, a BatchNormalization layer, and an ELU activation layer; (c) The illustrative process of depthwise Convolution. In EEGNeX, it acts as a frequency spatial learner/filter. (d) Visualization of dilation mechanism: the kernel scans the convolution layer with spaces in-between to increase the kernel receptive field.}
	\label{fig:4.jpg}
\end{figure}

This step focuses on the second part of the model, which consists of a combination of a depthwise and a separable convolution. Depthwise convolutions, initially proposed for efficient parameter usage in large image datasets training \citep{Howard.2017}, have two main advantages: 1) Its built-in mechanism acts as a cross-channel, frequency spatial learner (as shown in Figure \ref{fig:4.jpg}, subfigure c), improving the global information extraction abilities, especially in multi-channel EEG datasets. 2) They reduce network parameters and complexity. Given that EEG decoding typically handles smaller datasets than those used in in conventional CNNs like ResNet and EfficientNet in the CV domain, our enhancement should focus on boosting the network's parameter space.

Separable convolution involves decomposing a convolution kernel into two smaller ones \citep{Howard.2017}. It consists of a depthwise convolution (cross-channels) and a pointwise convolution (1*1 convolution to combine the outputs of the depthwise convolution). Since the channel dimension is compressed to one from the previous process, the separable convolution becomes redundant. In order to reinforce the temporal features learning ability in each extracted feature map and also to enhance the network's parameter space, we drop the separable convolution and add an extra Conv2D block with a kernel size of 16 and output filters as 16 (8 $\rightarrow$ 16: depthwise convolutions with a depth multiplier ratio of two).

\textbf{Step 3. Inverted bottleneck}

Following the proved design recommendation of advanced CNN architectures \citep{Tan.2019,Liu.2022,Sandler.2018}, we modified the network blocks (five blocks based on previous modifications) to have an inverted bottleneck structure with an expansion ratio of 4. The final output filters of CNN blocks are 8*32*64*32*8 (32 $\rightarrow$ 64: depthwise convolutions with a depth multiplier ratio of two).

\textbf{Step 4. Network dilation with fewer activation functions}

This step focuses on enhancing the third part of the network and reinforcing its temporal feature learning ability. Instead of adding sequential NNs such as causal padding or TCN block, a different approach is taken to avoid further increasing the complexity. Dilations is added to both Conv2D blocks in the third part of the network, with values of 1*2 and 1*4, respectively, to capture overall temporal features. Finally, following the suggestions in \citep{Liu.2022}, the activation functions between blocks are reduced. Now, each part of the network uses a single ELU activation.

All combined steps composite the EEGNeX. 

\textbf{Closing remarks}

We propose a new architecture of the network design after making several micro adjustments. Instead of the successful changes mentioned above, some attempts that failed to improve performance include:
\begin{enumerate}
    \item Deeper networks: Simply stacking more convolution layers often results in convergence difficulties during the training phase and poorer performance \citep{He.2016}. EEG datasets are typically smaller compared to image dataset, so a deep network would encounter the overfitting issue earlier. In fact, we realized that focusing on proper information extraction process by incorporating domain knowledge is more beneficial for the model performance than adding extra layers. For example, EEGNet-8.2, compared to other models, has fewer trainable parameters than other benchmark models, but it has a dominant performance advantage.
    \item Fewer normalization layers: Although removing some normalization layers in modern CNN-based structures in the computer vision domain results in some improvements, this tactic was not effective in our domain, potentially due to the shallow network design.
    \item Reordering different parts/layers: We experimented with various combinations of parts’ orderings based on EEGNet’s structure and tried attaching additional CNN-/RNN-based layer variations to the model. No improvements were observed. Instead, extra dilation with padding on the third part of the model helps to capture the global/temporal features efficiently.
    \item Attention block: We implemented a common attention structure: convolutional block attention module (CBAM) \citep{Woo.2018} between parts for channel and spatial attention. No improvements were found.
\end{enumerate}

Finally, we present a pure convolution-based network, EEGNeX-8,32, designed in line with modern CNN design practices. To conclude the design pillars of EEGNeX-8,32: we incorporate the local spatial and temporal information extraction ability by the DepthwiseConv2D, and enhances these features globally through the use of dilated Conv2D with padding. The network also utilizes an inverted bottleneck structure to improve its representation learning ability, while using fewer activation functions to avoid accuracy deterioration from multiple non-linear transformations. The results show that EEGNeX-8,32 achieves a significant improvement in accuracy on the four datasets compared to EEGNet-8,2. The complete structure and parameter description of EEGNeX-8,32 is presented in Figure \ref{fig:4.jpg} and Table \ref{tab:table3}.

\begin{table}
\resizebox{\textwidth}{!}{
\centering
\arrayrulecolor[rgb]{0.498,0.498,0.498}
\begin{tabular}{lllllll} 
\arrayrulecolor{black}\hline
\textbf{Block}      & \textbf{Layer}  & \textbf{\# filters} & \textbf{size} & \textbf{\# params}    & \textbf{Output}         & \textbf{Options}                                                                                                                 \\ 
\arrayrulecolor[rgb]{0.498,0.498,0.498}\hline
\textbf{1}          & Input           &                     &               &                       & (C, T)                  &                                                                                                                                  \\ 
\hline
                    & Reshape         &                     &               &                       & (1, C, T)               &                                                                                                                                  \\ 
\arrayrulecolor{black}\hline
                    & Conv2D          & \textit{$F_1$}         & (1, 32)       & \textit{32* $F_1$}       & (\textit{$F_1$}, C, T)     & use\_bias = False,
  padding='same'                                                                                              \\ 
\arrayrulecolor[rgb]{0.498,0.498,0.498}\hline
                    & BatchNorm       &                     &               & \textit{2*T}          & (\textit{$F_1$}, C, T)     &                                                                                                                                  \\ 
\arrayrulecolor{black}\hline
\textbf{2}          & Conv2D          & \textit{$F_1$*4}       & (1, 32)       & \textit{32* $F_1$*16}    & (\textit{$F_1$}*4, C, T)   & use\_bias = False,
  padding='same'                                                                                              \\ 
\arrayrulecolor[rgb]{0.498,0.498,0.498}\hline
                    & BatchNorm       &                     &               & \textit{2*T}          & (\textit{$F_1$}*4, C, T)   &                                                                                                                                  \\ 
\arrayrulecolor{black}\hline
                    & Activation      &                     &               &                       & (\textit{$F_1$}*4, C, T)   & ELU                                                                                                                              \\ 
\arrayrulecolor[rgb]{0.498,0.498,0.498}\hline
\textbf{3}          & DepthwiseConv2D & \textit{$F_1$*4*D}     & (C, 1)        & \textit{2*T}          & (\textit{$F_1$}*8, 1, T)   & \begin{tabular}[c]{@{}l@{}}depth\_multiplier=\textit{D}, use\_bias = False,  \\depthwise\_constraint=max\_norm(1.)\end{tabular}  \\ 
\arrayrulecolor{black}\hline
                    & BatchNorm       &                     &               & \textit{2*T}          & (\textit{$F_1$}*8, 1, T)   &                                                                                                                                  \\ 
\arrayrulecolor[rgb]{0.498,0.498,0.498}\hline
                    & Activation      &                     &               &                       & (\textit{$F_1$}*8, 1, T)   & ELU                                                                                                                              \\ 
\arrayrulecolor{black}\hline
\textbf{~}          & AvgPool2D       & \textit{~}          & (1, 4)        & \textit{~}            & (\textit{$F_1$}*8, 1, T/4) & ~                                                                                                                                \\ 
\arrayrulecolor[rgb]{0.498,0.498,0.498}\hline
                    & Dropout         &                     &               &                       & (\textit{$F_1$}*8, 1, T/4) & rate = 0.5                                                                                                                       \\ 
\arrayrulecolor{black}\hline
\textbf{4}          & Conv2D          & \textit{$F_1$*4*D}     & (1, 16)       & \textit{16* $F_1$*128/4} & (\textit{$F_1$}*8, 1, T/4) & \begin{tabular}[c]{@{}l@{}}use\_bias = False, padding='same',  \\dilation\_rate=(1, 2)\end{tabular}                              \\ 
\arrayrulecolor[rgb]{0.498,0.498,0.498}\hline
                    & BatchNorm       &                     &               & \textit{2*T/4}        & (\textit{$F_1$}*8, 1, T/4) &                                                                                                                                  \\ 
\arrayrulecolor{black}\hline
\textbf{5}          & Conv2D          & \textit{$F_1$}         & (1, 16)       & \textit{16* $F_1$*16/4}  & (\textit{$F_1$}, 1, T/4)   & \begin{tabular}[c]{@{}l@{}}use\_bias = False, padding='same',  \\dilation\_rate=(1, 4)\end{tabular}                              \\ 
\arrayrulecolor[rgb]{0.498,0.498,0.498}\hline
                    & BatchNorm       & \textit{~}          & ~             & \textit{2*T/4}        & (\textit{$F_1$}, 1, T/4)   & ~                                                                                                                                \\ 
\arrayrulecolor{black}\hline
                    & Activation      & \textit{~}          & ~             & \textit{~}            & (\textit{$F_1$}, 1, T/4)   & ELU                                                                                                                              \\ 
\arrayrulecolor[rgb]{0.498,0.498,0.498}\hline
                    & AvgPool2D       & \textit{~}          & (1, 8)        & \textit{~}            & (\textit{$F_1$}, 1, T/32)  & ~                                                                                                                                \\ 
\arrayrulecolor{black}\hline
                    & Dropout         &                     &               &                       & (\textit{$F_1$}, 1, T/32)  & rate = 0.5                                                                                                                       \\ 
\arrayrulecolor[rgb]{0.498,0.498,0.498}\hline
                    & Flatten         &                     &               & \textit{$F_1$*T/32}      & (\textit{$F_1$}*T/32)      &                                                                                                                                  \\ 
\arrayrulecolor{black}\hline
\textbf{Classifier} & Dense           &                     &               &                       & N                       & kernel\_constraint=max\_norm(0.25)                                                                                               \\ 
\arrayrulecolor[rgb]{0.498,0.498,0.498}\hline
                    & Activation      &                     &               &                       & N                       & softmax                                                                                                                          \\
\hline
\end{tabular}}
\caption{EEGNeX-8,32 architecture, where \textit{C}= number of channels,  \textit{T}= number of timesteps, \textit{$F_1$ }= number of temporal filters (8), \textit{D}= depth multiplier in DepthwiseConv2D (2), and \textit{N}= number of classes, respectively. }
\label{tab:table3}
\vspace{-0.5cm}
\end{table}

\section{Results: Comparison with existing methods}\label{Sec5}
\subsection{Model Complexity}

A critical aspect	of evaluating model performance or complexity often refers to the computational cost or time required to train the model and make predictions. It's usually associated with the following two factors: 
\begin{itemize}
    \item \textit{Number of parameters} in a model also contributes to its complexity. More parameters can potentially increase the model's capacity to fit the training data. However, more parameters also increase the risk of overfitting and demand more computational resources. The number of layers
    \item \textit{Depth of the network} in a model is one of the factors that contribute to its complexity. More layers allow the model to learn more complex representations, but it also makes the model more prone to overfitting if not properly regularized.
\end{itemize}

In the pursuit of evaluating both model performance and time complexity, we have taken into consideration various characteristics of the model architectures used for EEG signal decoding, including EEGNet, EEG-Inception, and EEG-TCNet, as presented in Table \ref{tab:table4}.

\definecolor{Black}{rgb}{0,0,0}
\definecolor{Gray}{rgb}{0.498,0.498,0.498}
\begin{table}[H]
\small
\centering
\begin{tblr}{
  width = \linewidth,
  colspec = {Q[175]Q[146]Q[167]Q[179]Q[154]Q[113]},
  column{even} = {c},
  column{3} = {c},
  column{5} = {c},
  hline{1} = {-}{},
  hline{2} = {-}{Black},
  hline{4} = {-}{Gray},
}
 & {\textbf{EEG-Inception}\\\textbf{\citep{Riyad.2020}}} & {\textbf{EEGNet-8,2}\\\textbf{\citep{Lawhern.2018}}} & {\textbf{EEG-TCNet}\\\textbf{\citep{Ingolfsson.2020}}} & {\textbf{EEG-ITNet}\\\textbf{\citep{Salami.2022}}} & \textbf{EEGNeX-8,32}\\
\textbf{Depth (\# layers)} & High
  (23) & Low (7) & High
  (17) & Very
  high (31) & Medium (10)\textbf{}\\
\textbf{Para. complexity} & Very
  high (\textasciitilde{}15k) & Low
  (\textasciitilde{}2k) & Medium
  (\textasciitilde{}5k) & Low
  (\textasciitilde{}3k) & High
  (\textasciitilde{}12k)
\end{tblr}
\caption{Comparison of the key factors of EEGNeX-8,32 with other end-to-end architectures.}
\label{tab:table4}
\vspace{-0.5cm}
\end{table}

Compared to other models, EEGNeX-8,32 maintains a reasonable balance between the two aspects of complexity, which promotes efficiency and reduces the risk of overfitting. For instance, although EEG-Inception has a very high parametric complexity ($\sim$15k), it also has a high depth (23 layers), which might increase the chances of overfitting and computational costs. On the other hand, despite having a low number of parameters ($\sim$2k), EEGNet-8,2 also has a lower depth (7 layers), potentially compromising its performance. EEGNeX-8,32 stands as a middle ground with a medium number of parameters ($\sim$12k) and high depth (10 layers). Through this balance of complexities, EEGNeX-8,32 aims to achieve optimal performance without excessive computational and memory demands, paving the way for practical applications in the field.

\subsection{BCI Competition IV Dataset 2a}

To validate the efficacy of EEGNeX, we first test its performance with similar advanced methods across the literature mentioned in Section \ref{Sec2} with their shared evaluation dataset: BCI Competition IV dataset 2a (SMR). Three extra existing methods were selected for performance evaluation: EEG-Inception, EEG-TCNet, and EEG-ITNet. We evaluated within-subject by using the same data processing implementation, resampling rate (125Hz), training strategy with the result described in \citep{Salami.2022}. Three EOG channels available in the dataset were excluded from our analysis. 

Table \ref{tab:table5} illustrate the comparison results with the test of significance, and the confidence interval for all candidate models. EEGNeX shows accuracy advance with statistical significance over EEGNet, EEG-Inception, and EEG-TCNet. The result suggests that EEGNeX (78.81\%) outperforms the other methods. It is also worth noting that the standard deviation values suggest that EEGNeX has a comparatively consistent performance.

\definecolor{Gray}{rgb}{0.498,0.498,0.498}
\definecolor{Black}{rgb}{0,0,0}
\begin{table}[h]
\small
\centering
\SetTblrInner{rowsep=0pt}
\begin{tblr}{
  width = \linewidth,
  colspec = {Q[94]Q[163]Q[185]Q[200]Q[169]Q[123]},
  column{even} = {c},
  column{3} = {c},
  column{5} = {c},
  hline{1} = {-}{},
  hline{2,15} = {-}{Gray},
  hline{11} = {-}{Black},
}
 & {\textbf{EEG-Inception}\\\textbf{\citep{Riyad.2020}}} & {\textbf{EEGNet-8,2}\\\textbf{\citep{Lawhern.2018}}} & {\textbf{EEG-TCNet}\\\textbf{\citep{Ingolfsson.2020}}} & {\textbf{EEG-ITNet}\\\textbf{\citep{Salami.2022}}} & \textbf{EEGNeX-8,32}\\
\textbf{S1} & 77.43 & 81.94 & 82.29 & 84.38 & \textbf{86.25}\\
\textbf{S2} & 54.51 & 56.94 & \textbf{64.24} & 62.85 & 60.71\\
\textbf{S3} & 82.99 & 90.62 & 88.89 & 89.93 & \textbf{93.38}\\
\textbf{S4} & \textbf{72.22} & 67.01 & 60.76 & 69.10 & 70.27\\
\textbf{S5} & 73.26 & 72.57 & 72.92 & \textbf{74.31} & 67.14\\
\textbf{S6} & 64.24 & 58.68 & 62.50 & 57.64 & \textbf{70.63}\\
\textbf{S7} & 82.64 & 76.04 & 83.33 & 88.54 & \textbf{88.84}\\
\textbf{S8} & 77.78 & 81.25 & 79.51 & 83.68 & \textbf{85.89}\\
\textbf{S9} & 76.36 & 78.12 & 76.39 & 80.21 & \textbf{86.16}\\
\textbf{Average} & 73.50 & 73.69 & 74.54 & 76.74 & \textbf{78.81}\\
\textbf{Std.} & 9.11 & 11.12 & 10.09 & 11.48 & 11.60\\
\textbf{CI*, 95\%} & {[}67.5, 79.5] & {[}66.5, 81.0] & {[}67.9, 81.1] & {[}69.2, 84.2] & \textbf{[71.2, 86.4]}\\
\textbf{p-value} & \textbf{0.011*} & \textbf{0.011*} & \textbf{0.015*} & 0.146 & \textbf{0.012*}
\end{tblr}
\caption{Performance results of advanced algorithms for within-subject evaluation of BCI Competition IV Dataset 2a in terms of classification accuracy. Star * corresponds to significant at level of 0.05; CI: Confidence Interval.}
\label{tab:table5}
\vspace{-0.75cm}
\end{table}

The confidence intervals (CI) presented in Table 4 demonstrate the range within which we can expect the true population parameter (in this case, classification accuracy) to fall 95\% of the time if the experiment were to be repeated. These intervals help us gauge the precision and reliability of our model's performance estimates. For instance, the 95\% CI for EEGNet-8,2 indicates that if the experiment were repeated multiple times, we could expect the true accuracy to fall within the range of 66.5 to 81.0, 95\% of the time. Similarly, for EEGNeX-8,32, the range would be 71.2 to 86.4.

The observed p-values in the table further confirm the statistical significance of the accuracy improvements exhibited by EEGNeX. A p-value of less than 0.05 (the typical threshold for statistical significance) indicates that the observed differences in performance are unlikely to have occurred by chance, further validating the effectiveness of EEGNeX.

\subsection{Mother of All BCI Benchmarks}
To exhaustively examine the efficacy of EEGNeX in a broader context, 11 open access, multi-classes MI datasets in MOABB \citep{Jayaram.2018} were applied. We benchmark EEGNeX and EEGNet on MOABB by comparing them to three other built-in common non-NN algorithm pipelines in BCIs. These algorithms are:

\begin{itemize}
\item CSP + LDA: where trail covariances estimated via maximum-likelihood with unregularized common spatial patterns (CSP). Features were log variance of the filters belonging to the six most diverging eigenvalues and then classified with linear discriminant analysis (LDA). 
\item TS + optSVM: where trial covariances estimated via oracle approximating shrinkage estimator (OAS) then projected into the Riemannian tangent space to obtain features and classified with a linear SVM with identical grid search. 
\item AM + optSVM: where features are the log-variance in each channel and then classified with a linear SVM with grid search.
\end{itemize}

\begin{figure}[h]
	\centering
	\includegraphics[width=16cm]{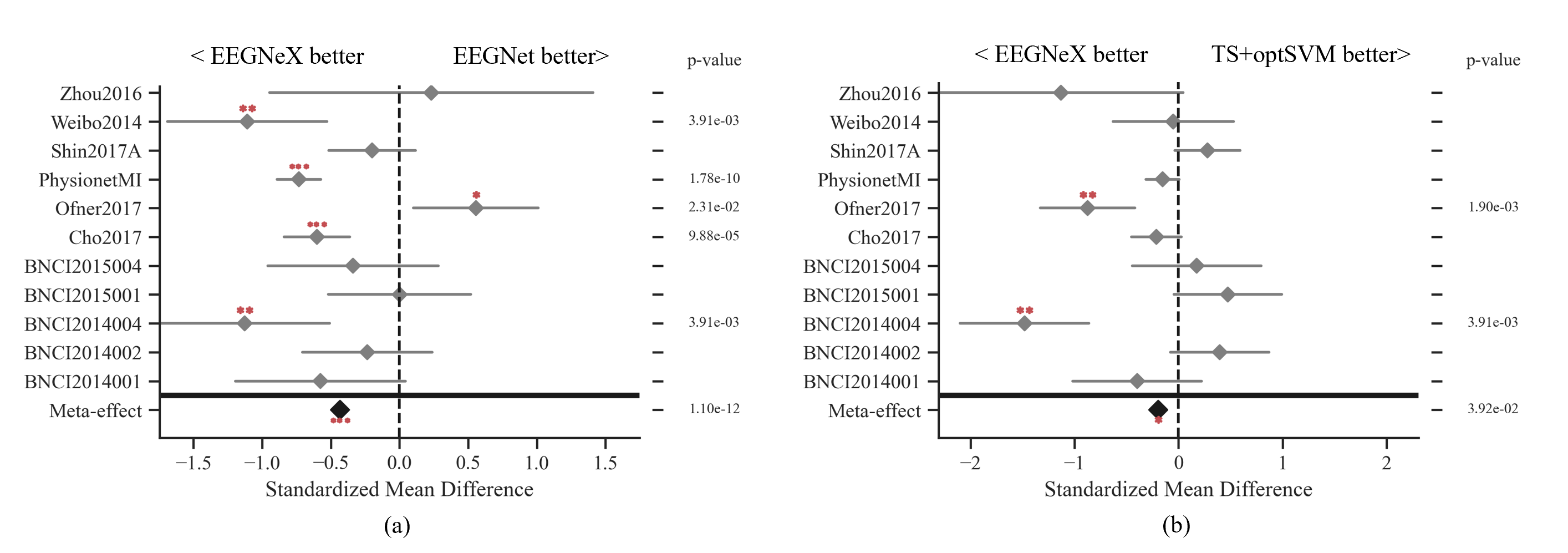}
	\caption{Meta-analysis comparing EEGNeX against EEGNet (a) and the state-of-the-art algorithm pipeline on MOABB: TS-optSVM (b). The result is evaluated by within-recoding-session accuracy. The effect sizes shown are standardized mean differences, with p-values corresponding to the one-tailed Wilcoxon signed-rank test for the hypothesis given at the top of the plot and 95\% interval denoted by the grey bar. Stars correspond to *** =p < 0.001, ** = p < 0.01, * = p < 0.05. The meta-effect is shown at the bottom of the plot. The overall trend shows that EEGNeX is on average better than EEGNet and TS-optSVM.}
	\label{fig:5.jpg}
\end{figure}

The evaluation was performed within-subject to minimize the impact of non-stationarity. It should be noted that some dataset collections within MOABB have small sample sets of data within-session, which increases the training difficulties of NN algorithms. The accuracy was evaluated using 5-fold cross-validation to validate the general information extraction and representation performance. It is worth noting that there is a “No-free-lunch” theorem \citep{wolpert1997no},  or “horses for courses” phenomenon exists in the algorithm selection problem, which states that no single algorithm can be deemed the best for all problems. Figure \ref{fig:5.jpg} presents the meta-analysis of the performance comparison between EEGNeX, EEGNet, and TS+optSVM, which is the state-of-the-art algorithm pipeline on MOABB. Across diverse datasets, the result validates the performance novelty of EEGNeX: it competes favorably with EEGNet methods (p < 0.001, Wilcoxon tests) and outperforms TS+optSVM (p < 0.05).

\begin{figure}[h]
	\centering
	\includegraphics[width=10cm]{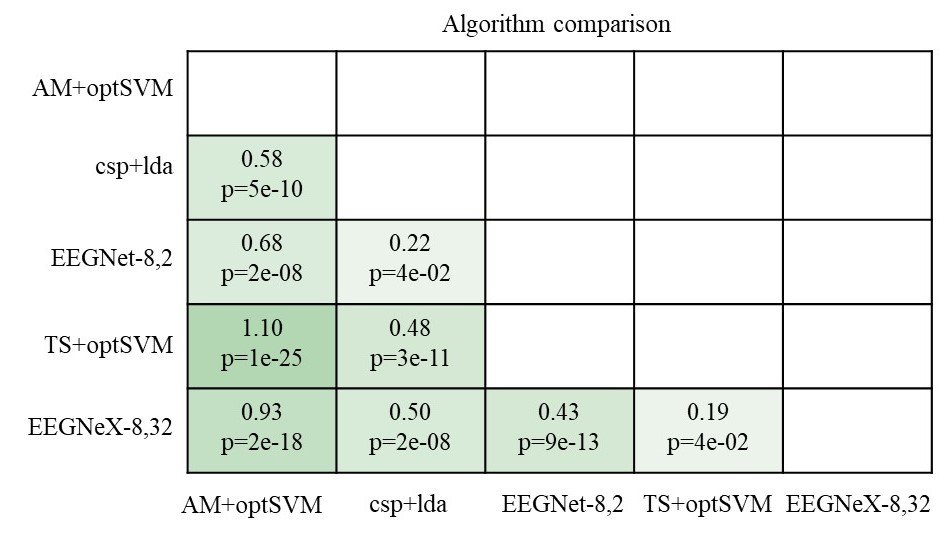}
	\caption{Ranking of algorithms in performance across all datasets. As all p-values are single-sided for clarity. The values correspond to the standardized mean difference of the algorithm in the y-axis minus that in the x-axis and the associated p-value. The comparison shows the meta-effect in case that the method on the vertical axis significantly outperforms the method on the horizontal axis, according to the one-tailed Wilcoxon signed-rank test.}
	\label{fig:6.jpg}
\end{figure}

Figure \ref{fig:6.jpg} summarizes the ranking of algorithms in performance with statistics generated across all 11 datasets in MOABB. The result underlines that EEGNeX generalized well across all datasets and outperformed other algorithm pipelines. We concluded that the general structure modifications to EEGNeX (increased layer depth, inverted bottleneck, dilation) and micro adjustment are universally valid for accuracy improvement in EEG-based BCI classification tasks.

In the end, we open-sourced code in Keras for easy, out-of-box reproduction to use in different research sources (\href{https://github.com/chenxiachan/EEGNeX}{(https://github.com/chenxiachan/EEGNeX)}). All models are coded in the same implementation format with input in the shape of (\textit{trials * channels * timesteps}), and one-hot-encoded output. We designed an experiment to run all models on the dataset for ten rounds with random seeds to record their accuracy performances for statistical tests.

\section{Discussion}\label{Sec6}

The thought in our brain, has been using mainly language to interact with the world for a long period. Just as Ludwig Josef Johann Wittgenstein \citep{Wittgenstein.2013} described:

\centerline{ }
\centerline{\textit{“The limits of my language mean the limits of my world.”}}

Brain-computer interface (BCI) research opens up a direct pathway for decoding brain activity, offering immense potential for creating new patterns of human-computer interaction (HCI). By the generalization power of NNs, we foresee that the effort invested in interpreting brain activities would benefit a broad branch of domains, from information \& communications technology to medical, from social science to design, and even revolutionize how we percept and understand reality. Take the design domain as an example; new BCI patterns would efficiently assist designers in expressing and interacting with the design work: their personal preferences, less expressible feelings, and subconscious perceptions can be captured, represented, and transferred more seamlessly via the brain’s electrical signals without explicitly formalizing in languages or actions. 

Starting with the EEG classification, this study aims to reduce the gap between the BCI domain research and efficient NN implementation. It offers a testbed for different network architectures comparison to encourage more interdisciplinary research participation. From this perspective, the construction of our open-source testbed remains primitive.

Our investigation focused on designing NNs effectively for decoding EEG signals. Despite basing our general EEGNeX model on established literature and past experiences, we noticed that different NN layers correspond to different numbers of trainable parameters during the benchmarking process. The number of trainable parameters in a NN indicates its capacity, i.e., the amount of information it can retain and employ to make predictions. However, deeper networks do not necessarily imply larger model capacity. The benchmarking results confirm that the efficiency of information extraction by the NN surpasses the importance of merely expanding model capacity. Selecting the right number of trainable parameters for a task involves a balance: the need for a high-capacity model must be weighed against the risk of overfitting. This balance can be influenced by factors such as the size and complexity of the input data, the availability of training examples, and the computational resources at the disposal of model training. The benchmarking of EEGNeX against various architectures revealed crucial insights. The model's efficiency in information extraction surpasses the importance of merely expanding the model's capacity. This emphasizes the judicious selection of trainable parameters to balance the need for a high-capacity model against the risk of overfitting. 

With respect to confidence intervals, they provide a measure of uncertainty around our mean accuracy estimates. These intervals underline the inherent variability in model performance across subjects and suggest a potential value in individualized model tuning. It's worth noting that the broad range of this confidence interval may be due to the relatively small sample size and the inherent variability of EEG signals. Future work with larger sample sizes may lead to narrower confidence intervals, thus improving the precision of the estimate. Although the Leave-One-Subject-Out (LOSO) cross-validation method was not utilized in this study, it could be an area for future exploration in cross-subject evaluation to gauge the performance of these models in different settings. 

The limitation of this study is: The intention of designing this benchmark is to include diverse NN architectures in the research of reliable EEG signal decoding paths. We referred to ConvNeXt to find the path shortcut for improved efficiency by the convolution inductive bias. Although the original ConvNeXt claims the performance advance in the CV domain, image classification tasks than Transformer, the contribution of different mechanisms in four pillars mentioned in Figure \ref{fig:2.jpg} is still worth investigating in the BCI domain. 

Aside from the performance advantages of CNN-based methods, we noted from our benchmark results that hybrid-model approaches yield accuracy improvements over single architectures, hinting at a more profound question of EEG representation. The rationale is that BCI signal data fundamentally combines the sequential format of NLP and the multi-channels from image data in computer vision. We have noticed various research attempts to utilize BCI data across both branches \citep{Hosseini.2020,Craik.2019}. In this context, the hybrid-model approach by deeply combining CNN and RNN architecture, the decent design of acquiring long-range global features in sequence by self-attention in Transformer, dimension reduction by encoder-decoder, and more NN techniques own potential for further improvement. Furthermore, our study and proposed EEGNeX aim for a method that is generalizable for EEG decoding without explicit feature extraction. A domain-knowledge embedding into the network structure, objective function, or feature engineering owns potential for further performance enhancement in a specific dataset. 

In terms of generalization evaluation, future research could test the decoding performance of EEGNeX using time-series data from other imaging modalities such as magnetoencephalography (MEG) and functional near-infrared spectroscopy (fNIRS). For the utility, it is worth noting that although the model is developed to address the challenges of EEG tasks, the benchmark methods and the improvement conducted in the EEGNeX architecture rarely incorporate domain-specific feature engineering knowledge. Broader test scenarios within the scope of biological signal decoding in BCI topics could also benefit from this study.

\section{Conclusion}\label{Sec7}

In this study, we conducted experiments on large-scale neural networks (NNs) to evaluate their performance in EEG-based brain-computer interface (BCI) classification tasks. This deep dive allowed us to delineate clear patterns regarding the efficacy and representation capabilities of different NN structures in EEG decoding. Our findings identified four pivotal pillars essential for architecting efficient neural networks tailored for EEG classification.

With these insights, we introduced EEGNeX – a pioneering, convolution-centric network model that leverages breakthroughs from contemporary neural network representation research. Beyond theoretical propositions, we rigorously evaluated EEGNeX across a spectrum of EEG classification challenges, affirming its consistent and universal effectiveness. Crucially, EEGNeX's standout performance was not tethered to any context-specific design paradigms, endowing it with an adaptable architecture. This adaptability empowers EEGNeX to synergize seamlessly with knowledge-driven and domain-centric feature engineering, permitting tailored applications across varied scenarios.

Through this study and the accompanying benchmarks, our objective has been to simplify and streamline the deployment of advanced neural models within the BCI research domain. We're optimistic that our endeavors will not only facilitate more straightforward model implementation but also catalyze future representation-centric explorations in this sphere.

\section{Acknowledgments including declarations}
This article does not contain any studies involving human participants or animals performed by any of the authors. All data used in this study are openly available datasets. The authors declare that they have no conflicts of interest.

\section{Appendix}\label{Apd}

\subsection{Benchmark model structures}
The benchmark model was designed with the following considerations:
\begin{enumerate}
    \item Three-Block Architecture: All basic models were constructed in a three-block manner to be comparable with EEGNet, a model with a three-block architecture with proven performance in EEG-based tasks. This three-block architecture was chosen because it would be suitable for reflecting the model performance in EEG-based tasks without worrying about the model capacity.
    \item Common Macro Design: All models followed a common macro design based on a large-scale literature review of NN. The design includes the following elements:

    \begin{itemize}

        \item Input layer: Each NN model starts with an input layer to receive the EEG data.
        \item Hidden layers: The hidden layers are designed to extract features from the EEG data. They consist of a certain type of network layer, batch normalization, and activation layers. In CNN-based models, a pooling layer is also included. Some layer variations are also made for comparison.
        \item Output layer: The hidden layers are followed by a fully connected dense layer with 100 neurons before the output. A dropout layer is attached to enhance the model's robustness.

    \end{itemize}

    \item Hybrid Model: A hybrid model was designed by attaching an RNN layer after the CNN blocks to test the model performance by combining spatial-temporal information extraction design. This design aims to capture both spatial and temporal information from the EEG data to improve the model's performance in EEG-based tasks.
\end{enumerate}
In summary, the benchmark model was designed to be comparable with previous research and reflect the performance of different NN models in EEG-based tasks. The design followed a common macro design based on a large-scale literature review and incorporated a hybrid model to test the performance of combining spatial-temporal information extraction.

\definecolor{Black}{rgb}{0,0,0}
\definecolor{Gray}{rgb}{0.498,0.498,0.498}
\begin{table}[H]
\small
\centering
\SetTblrInner{rowsep=0pt}
\begin{tblr}{
  width = \linewidth,
  colspec = {Q[137]Q[102]Q[87]Q[54]Q[556]},
  cell{1}{2} = {c=2}{0.189\linewidth},
  hlines,
  vline{2-3,5} = {1}{Black},
  vline{2-5} = {2-9}{Black},
  hline{2-3,5,7,9-10} = {5}{Gray},
  hline{4,6,8} = {5}{Black},
}
\textbf{Single\_GRU} & \textbf{Single\_LSTM} &  &  & \\
\textbf{Layer} & \textbf{Layer} & \textbf{\# units} & \textbf{size} & \textbf{Options}\\
Input & Input &  & / & \\
GRU & LSTM & 100 & / & return\_sequences=True,
  kernel\_regularizer=l2(0.0001)\\
GRU & LSTM & 100 & / & return\_sequences=True, kernel\_regularizer=l2(0.0001)\\
GRU & LSTM & 100 & / & kernel\_regularizer=l2(0.0001)\\
Dropout & Dropout &  & / & rate = 0.5\\
Dense & Dense & 100 & / & activation='elu'\\
Dense & Dense &  & / & activation='softmax'
\end{tblr}
\caption{The architecture of GRU and LSTM; They both share the same general structure but with different RNN layer.}
\label{tab:table6}
\vspace{-0.5cm}
\end{table}

\definecolor{Black}{rgb}{0,0,0}
\definecolor{Gray}{rgb}{0.498,0.498,0.498}
\begin{table}[H]
\small
\centering
\SetTblrInner{rowsep=0pt}
\begin{tblr}{
  width = \linewidth,
  colspec = {Q[133]Q[79]Q[48]Q[87]Q[152]Q[156]Q[279]},
  column{even} = {c},
  column{3} = {c},
  column{5} = {c},
  column{7} = {c},
  cell{6}{4} = {c=4}{0.674\linewidth},
  cell{8}{4} = {c=4}{0.674\linewidth},
  cell{9}{4} = {c=4}{0.674\linewidth},
  cell{12}{4} = {c=4}{0.674\linewidth},
  cell{13}{4} = {c=4}{0.674\linewidth},
  cell{14}{4} = {c=4}{0.674\linewidth},
  cell{16}{4} = {c=4}{0.674\linewidth},
  cell{17}{4} = {c=4}{0.674\linewidth},
  hlines,
  vline{2-7} = {1-5,7,10-11,15}{Black},
  vline{2-4} = {6,8-9,12-14,16-17}{Black},
  hline{2-3,5,11} = {7}{Gray},
  hline{4,6,8,12,16} = {7}{Black},
  hline{7,9,13,15,17-18} = {4-7}{Gray},
  hline{10,14} = {4-7}{Black},
}
 &  &  & \textbf{1D\_CNN} & \textbf{1D\_CNN\_Dilated} & \textbf{1D\_CNN\_Causal} & \textbf{1D\_CNN\_CausalDilated}\\
\textbf{Layer} & \textbf{\# filters} & \textbf{size} & \textbf{Options} & \textbf{Options} & \textbf{Options} & \textbf{Options}\\
Input &  &  &  &  &  & \\
Conv1D & 64 & 3 & / & / & padding='causal' & padding='causal'\\
BatchNorm &  &  &  &  &  & \\
Activation &  &  & elu &  &  & \\
Conv1D & 64 & 3 & / & dilation\_rate=2 & padding='causal' & padding='causal', dilation\_rate=2\\
BatchNorm &  &  &  &  &  & \\
Activation &  &  & elu &  &  & \\
Conv1D & 64 & 3 & / & / & padding='causal' & padding='causal'\\
BatchNorm &  &  &  &  &  & \\
Activation &  &  & elu &  &  & \\
Dropout &  &  & rate = 0.5 &  &  & \\
MaxPooling1D &  &  & pool\_size=2 &  &  & \\
Flatten &  &  &  &  &  & \\
Dense & 100 &  & activation='elu' &  &  & \\
Dense &  &  & activation='softmax' &  &  & 
\end{tblr}
\caption{The architecture of 1D CNN models; They share the same general structure but with different layer parameter options.}
\label{tab:table7}
\vspace{-0.5cm}
\end{table}

\definecolor{Black}{rgb}{0,0,0}
\definecolor{Gray}{rgb}{0.498,0.498,0.498}
\begin{table}[H]
\small
\centering
\SetTblrInner{rowsep=0pt}
\begin{tblr}{
  width = \linewidth,
  colspec = {Q[235]Q[138]Q[90]Q[162]Q[302]},
  column{even} = {c},
  column{3} = {c},
  column{5} = {c},
  cell{6}{4} = {c=2}{0.463\linewidth},
  cell{9}{4} = {c=2}{0.463\linewidth},
  cell{12}{4} = {c=2}{0.463\linewidth},
  cell{13}{4} = {c=2}{0.463\linewidth},
  cell{14}{4} = {c=2}{0.463\linewidth},
  cell{16}{4} = {c=2}{0.463\linewidth},
  cell{17}{4} = {c=2}{0.463\linewidth},
  hlines,
  vline{2-5} = {1-5,7-8,10-11,15}{Black},
  vline{2-4} = {6,9,12-14,16-17}{Black},
  hline{2-3,5,9,11} = {5}{Gray},
  hline{4,6,8,12,16} = {5}{Black},
  hline{7,13,15,17-18} = {4-5}{Gray},
  hline{10,14} = {4-5}{Black},
}
 &  &  & \textbf{2D\_CNN} & \textbf{2D\_CNN\_Dilated}\\
\textbf{Layer} & \textbf{\# filters} & \textbf{size} & \textbf{Options} & \textbf{Options}\\
Input &  &  &  & \\
Conv2D & 64 & (1, 3) &  & \\
BatchNorm &  &  &  & \\
Activation &  &  & elu & \\
Conv2D & 64 & (1, 3) & / & dilation\_rate=2\\
BatchNorm &  &  &  & \\
Activation &  &  & elu & \\
Conv2D & 64 & (1, 3) &  & \\
BatchNorm &  &  &  & \\
Activation &  &  & elu & \\
Dropout &  &  & rate = 0.5 & \\
AvgPooling2D &  &  & pool\_size=2,
  padding='same' & \\
Flatten &  &  &  & \\
Dense & 100 &  & activation='elu' & \\
Dense &  &  & activation='softmax' & 
\end{tblr}
\caption{The architecture of 2D CNN models; They share the same general structure but with different layer parameter setting options.}
\label{tab:table8}
\vspace{-0.5cm}
\end{table}

\definecolor{Black}{rgb}{0,0,0}
\definecolor{Gray}{rgb}{0.498,0.498,0.498}
\begin{table}[H]
\small
\centering
\SetTblrInner{rowsep=0pt}
\begin{tblr}{
  width = \linewidth,
  colspec = {Q[231]Q[223]Q[98]Q[69]Q[317]},
  column{3} = {c},
  column{4} = {c},
  hlines,
  vline{2-5} = {-}{Black},
  hline{2-3,5,7,9,11,13,15,17-18} = {5}{Gray},
  hline{4,6,8,10,12,14,16} = {5}{Black},
}
\textbf{2D\_CNN\_Depthwise} & \textbf{2D\_CNN\_Separable} &  &  & \\
\textbf{Layer} & \textbf{Layer} & \textbf{\# filters} & \textbf{size} & \textbf{Options}\\
Input & Input &  &  & \\
DepthwiseConv2D & SeparableConv2D & 64 & (1, 3) & \\
BatchNorm & BatchNorm &  &  & \\
Activation & Activation &  &  & elu\\
SeparableConv2D & SeparableConv2D & 64 & (1,
  3) & \\
BatchNorm & BatchNorm &  &  & \\
Activation & Activation &  &  & elu\\
SeparableConv2D & SeparableConv2D & 64 & (1, 3) & \\
BatchNorm & BatchNorm &  &  & \\
Activation & Activation &  &  & elu\\
Dropout & Dropout &  &  & rate = 0.5\\
AvgPooling2D & AvgPooling2D &  &  & pool\_size=2,
  padding='same'\\
Flatten & Flatten &  &  & \\
Dense & Dense & 100 &  & activation='elu'\\
Dense & Dense &  &  & activation='softmax'
\end{tblr}
\caption{The architecture of Depthwise and Separable 2D CNN.}
\label{tab:table9}
\vspace{-0.5cm}
\end{table}

\definecolor{Black}{rgb}{0,0,0}
\definecolor{Gray}{rgb}{0.498,0.498,0.498}
\begin{table}[H]
\renewcommand\arraystretch{0.8}
\small
\centering
\SetTblrInner{rowsep=0pt}
\begin{tblr}{
  width = \linewidth,
  colspec = {Q[125]Q[137]Q[125]Q[137]Q[81]Q[79]Q[254]},
  column{1} = {c},
  column{even} = {c},
  column{3} = {c},
  column{5} = {c},
  cell{5}{1} = {c=4}{0.524\linewidth,c},
  cell{6}{1} = {c=4}{0.524\linewidth,c},
  cell{8}{1} = {c=4}{0.524\linewidth,c},
  cell{9}{1} = {c=4}{0.524\linewidth,c},
  cell{11}{1} = {c=4}{0.524\linewidth,c},
  cell{12}{1} = {c=4}{0.524\linewidth,c},
  cell{13}{1} = {c=4}{0.524\linewidth,c},
  cell{14}{1} = {c=2}{0.262\linewidth,c},
  cell{14}{3} = {c=2}{0.262\linewidth},
  cell{15}{1} = {c=4}{0.524\linewidth,c},
  cell{17}{1} = {c=4}{0.524\linewidth,c},
  cell{18}{1} = {c=4}{0.524\linewidth,c},
  cell{19}{1} = {c=4}{0.524\linewidth,c},
  hlines,
  vline{2-7} = {1-4,7,10,16}{Black},
  vline{2,6-7} = {5-6,8-9,11-13,15,17-19}{Black},
  vline{2,4,6-7} = {14}{Black},
  hline{2-3,5,7,9,11,13,15,17,19-20} = {7}{Gray},
  hline{4,6,8,10,12,14,16,18} = {7}{Black},
}
\textbf{CNN\_GRU1D} & \textbf{CNN\_LSTM1D} & \textbf{CNN\_GRU2D} & \textbf{CNN\_LSTM2D} &  &  & \\
\textbf{Layer} & \textbf{Layer} & \textbf{Layer} & \textbf{Layer} & \textbf{\# filters} & \textbf{size} & \textbf{Options}\\
Input & Input & Input & Input &  &  & TimeDistributed\\
Conv1D & Conv1D & Conv2D & Conv2D & 64 & 3/(1, 3) & TimeDistributed\\
BatchNorm &  &  &  &  &  & \\
Activation &  &  &  &  &  & elu\\
Conv1D & Conv1D & Conv2D & Conv2D & 64 & 3/(1, 3) & TimeDistributed\\
BatchNorm &  &  &  &  &  & \\
Activation &  &  &  &  &  & elu\\
Conv1D & Conv1D & Conv2D & Conv2D & 64 & 3/(1, 3) & TimeDistributed\\
BatchNorm &  &  &  &  &  & \\
Activation &  &  &  &  &  & elu\\
Dropout &  &  &  &  &  & TimeDistributed,
  rate = 0.5\\
MaxPooling1D &  & AvgPooling2D &  &  &  & TimeDistributed,
  pool\_size=2\\
Flatten &  &  &  &  &  & TimeDistributed\\
GRU & LSTM & GRU & LSTM & 480 &  & kernel\_regularizer=l2(0.0001)\\
Dropout &  &  &  &  &  & rate
  = 0.5\\
Dense &  &  &  & 100 &  & activation='elu'\\
Dense &  &  &  &  &  & activation='softmax'
\end{tblr}
\caption{The architecture of hybrid-models; They both share the same general structure but with different RNN layer.}
\label{tab:table10}
\vspace{-0.5cm}
\end{table}

\subsection{Description of datasets}

\textbf{Event-related potential (ERP) – MindBigData}

A single subject, 10-class ERP dataset \citep{Vivancos.}. This open database contains four different devices, 1,207,293 brain signals of 2 seconds long each. The signal records the imagine of presented digit images after exposing a subject to the visual stimulus of the MNIST dataset (from 0 to 9, -1 represents noise), resulting in similar phase synchrony among multiple channels for all classes. In this study, Emotive EPOC device data were selected. The raw EEG signals were recorded at a sample rate of 128 Hz; there are around 6500 trials for each digit image, with each trial containing 256 timesteps for 14 channels. We conducted a simple and common data preprocessing strategy, including: Noise events removal (-1); A combination of Butterworth lowpass filter (with a cutoff frequency of 63Hz) and a notch filter at 50Hz is applied on all trials; The first 32 timesteps (250ms) for each trial are trimmed to avoid noises caused by sensor power on. 

\textbf{Motor Imagery (MI) - OpenMBI Data}

A 2-class MI data from Korea University EEG dataset \citep{Lee.2019}. The dataset contains 2-class EEG data from 54 healthy subjects with MI of left/right-hand classes in a total of 100 trials for each session in the length of 4s \citep{Lee.2019}. The original data is recorded at 1000Hz using 62 electrodes. As suggested by the original work, we selected 20 channels (FC-5/3/1/2/4/6, C-5/3/1/z/2/4/6, and CP-5/3/1/z/2/4/6) for the classification task. The EEG data is filtered with a notch of 1-40Hz and down-sampled at 128Hz.

\textbf{Sensory Motor Rhythm (SMR) - BCIC-IV-2A Data}

A 4-class SMR data from BCI Competition IV Dataset 2A \citep{Tangermann.2012}. This dataset consists of 22 EEG channels and 3 EOG channels from 9 subjects with the task of 4-class motor imagery (left hand, right hand, feet, tongue) classification. The sampling rate is 250 Hz with 0.5-100Hz notch filtered. In our analysis, we used all EEG channel signals with the entire trial. 

\textbf{Feedback Error-Related Negativity (ERN) - BCI challenge Data}

A 2-class ERP data from BCI Challenge hosted by Kaggle \citep{Margaux.2012}. Detecting the ERN feedback helps to improve the performance of the P300 speller in the application. The dataset is used in the “BCI Challenge” to determine whether the P300 feedback is correct (2-class classification task), hosted by Kaggle \citep{Margaux.2012}, consists of 26 healthy participants in 56 passive Ag/AgCl EEG sensors. The data is originally recorded at 600Hz, we used a 1-40Hz notch filter with 128Hz down-sampling for analysis.

\bibliographystyle{unsrtnat}
\bibliography{references} 

\end{document}